# NP-completeness of Certain Sub-classes of the Syndrome Decoding Problem

Matthieu Finiasz

*Abstract*—The problem of Syndrome Decoding was proven to be NP-complete in 1978 and, since then, quite a few cryptographic applications have had their security rely on the (provable) difficulty of solving some instances of it. However, in most cases, the instances to be solved follow some specific constraint: the target weight is a function of the dimension and length of the code. In these cases, is the Syndrome Decoding problem still NP-complete? This is the question that this article intends to answer.

*Index Terms*—Syndrome Decoding, NP-completeness, Goppa codes.

## I. INTRODUCTION

BACK in 1978, Berlekamp, McEliece and van Tilborg [1] have proven that the coding theory problems of COSET WEIGHTS and SUBSPACE WEIGHTS are NP-complete. These problems, usually known by the name of SYNDROME DECODING problem (SD), are central in code-based cryptography. For instance, the security of the public key cryptosystems of McEliece or Niederreiter directly relies on the difficulty of solving an instance of SD. However, these instances are not generic instances: the need for a trap (the secret key), allowing to decrypt, adds some constraints on the instances to solve. This allows to define a sub-class of the SD problem. In this article, I prove that this sub-class is also an NP-complete problem and propose to generalize this result to some other sub-classes of the SD problem.

### A. General Context

Let $C$ be a $[n,k]$ binary linear code defined by its parity check matrix $\mathcal{H}$ of size $n \times r$ where $r = n-k$. The syndrome of a word $y$ is $\mathcal{S} = y\mathcal{H}$. An element of $C$ is called a codeword, and is a word of null syndrome. The weight of a word $y$ denotes the Hamming weight of the word, that is, the number of non-zero bits in $y$. Knowing this, we can state the two NP-complete problems of [1].

COSET WEIGHT:
 *Input:* a binary matrix $\mathcal{H}$, a syndrome $\mathcal{S}$ and a non negative integer $w$.
 *Property:* there exists a word $y$ of weight $\leq w$ such that $y\mathcal{H} = \mathcal{S}$.

SUBSPACE WEIGHT:
 *Input:* a binary matrix $\mathcal{H}$ and a non negative integer $w$.
 *Property:* there exists a codeword $x$ of weight $w$ such that $x\mathcal{H} = 0$.

The first of these two problems corresponds exactly to the problem of decoding up to a given distance: if one can solve this problem, he can decode up to $w$ errors in the code $C$ defined by the parity check matrix $\mathcal{H}$.

The second one is nearly a sub-class of the first problem, but not exactly: the syndrome is set to 0, but the weight must also be exactly equal to $w$, so as to avoid having the null word as a trivial solution.

### B. The Original NP-completeness Proof

To prove that the two previous problems are NP-complete, Berlekamp, McEliece and van Tilborg build reductions from the problem of THREE DIMENSIONAL MATCHING.

THREE DIMENSIONAL MATCHING:
 *Input:* a subset $U \subseteq T \times T \times T$, where $T$ is a finite set.
 *Property:* there exists a set $W \subseteq U$ such that $|W| = |T|$, and no two elements of $W$ agree in any of there three coordinates.

This new problem might not seem closely related to the previous decoding problems, but a short example makes the relation obvious. Let $T = \{1,2,3\}$ and $|U| = 5$ with:

$$\begin{aligned} U_1 &= (1,2,2) \\ U_2 &= (2,2,3) \\ U_3 &= (1,3,2) \\ U_4 &= (2,1,3) \\ U_5 &= (3,3,1) \end{aligned}$$

One can see that $U_1$, $U_4$, and $U_5$ verify the property. However, as soon as we remove $U_1$ from $U$, no solution exists. To relate this problem to SD, we introduce the incidence matrix $\mathcal{A}$ defined as follows:

$$\mathcal{A} = \begin{array}{c} \\ U_1 \\ U_2 \\ U_3 \\ U_4 \\ U_5 \end{array} \begin{array}{|ccc|ccc|ccc|} \multicolumn{3}{c}{1\ \ 2\ \ 3} & \multicolumn{3}{c}{1\ \ 2\ \ 3} & \multicolumn{3}{c}{1\ \ 2\ \ 3} \\ \hline 1 & 0 & 0 & 0 & 1 & 0 & 0 & 1 & 0 \\ 0 & 1 & 0 & 0 & 1 & 0 & 0 & 0 & 1 \\ 1 & 0 & 0 & 0 & 0 & 1 & 0 & 1 & 0 \\ 0 & 1 & 0 & 1 & 0 & 0 & 0 & 0 & 1 \\ 0 & 0 & 1 & 0 & 0 & 1 & 1 & 0 & 0 \\ \hline \end{array}$$

This binary matrix $\mathcal{A}$ of size $|U| \times 3|T|$ contains three 1 per line, one for each coordinate, in the column corresponding to the correct element of $T$. With this new representation of the set $U$, finding a valid solution to the THREE DIMENSIONAL MATCHING problem corresponds to finding a set of $|T|$ rows of $\mathcal{A}$ summing to the all 1 vector.

If an algorithm is able to solve any instance of the COSET WEIGHT problem, it is enough to feed it with the parameters $\mathcal{H} = \mathcal{A}$, $\mathcal{S} = (1,\ldots,1)$ and $w = |T|$ to solve any given instance of THREE DIMENSIONAL MATCHING. This proves that COSET WEIGHT is NP-complete.

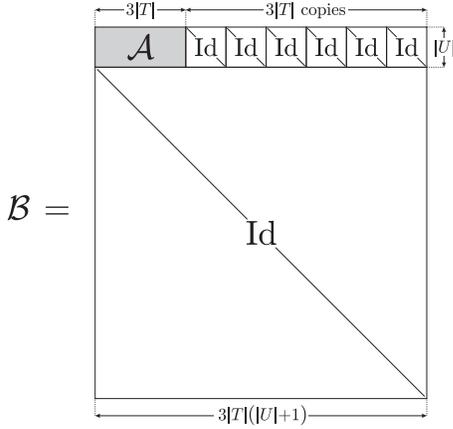

Fig. 1. Matrix used to reduce THREE DIMENSIONAL MATCHING to SUBSPACE WEIGHT.

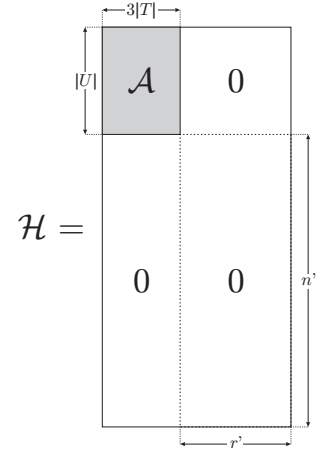

Fig. 2. Matrix used to reduce THREE DIMENSIONAL MATCHING to GOPPA PARAMETERIZED SYNDROME DECODING.

For the second reduction, Berlekamp, McEliece, and van Tilborg propose to build the matrix $\mathcal{B}$ represented on Fig. 1, still based on the incidence matrix $\mathcal{A}$. If an algorithm is able to solve any instance of SUBSPACE WEIGHT, we can use it with parameters $\mathcal{H} = \mathcal{B}$ and $w = 3|T|^2 + 4|T|$ to solve the instance of THREE DIMENSIONAL MATCHING represented by matrix $\mathcal{A}$. For instance, any null sum of $w$ rows of $\mathcal{B}$ containing $\alpha$ lines among the $|U|$ top ones, will necessarily contain between $3\alpha|T|$ and $3\alpha|T| + 3\alpha$ lines in the bottom part (this is because each of the first $|U|$ lines has weight $3|T| + 3$, and $3|T|$ in positions which cannot overlap). Thus, any solution to SUBSPACE WEIGHT with $w = 3|T|^2 + 4|T|$ must verify $\alpha = |T|$ and has a weight of exactly $3|T|^2 + 3|T|$ in the bottom part with the $|T|$ rows of $\mathcal{A}$ not overlapping. It is also a solution to THREE DIMENSIONAL MATCHING.

## II. SOME NEW PROBLEMS

### A. Using Goppa Codes

In the McEliece or Niederreiter cryptosystems, the matrix of a Goppa code is scrambled and used as the public key, the secret key is the scrambling matrices that allow to use the standard Goppa decoding algorithm. The security of these construction thus relies on the hardness of two problems: recovering the structure of the Goppa code by unscrambling the public matrix, and decoding as in a random code. However, this code is not completely random: even if the matrix is not distinguishable from a random binary matrix of the same size, the decoding problem that has to be solved uses parameters which are those of a Goppa code. This means that the code has a length $n = 2^m$ and a dimension $k = n - mt$, where $t$ is the correction capacity of the code. Decoding requires to solve an instance of SD (the COSET WEIGHT problem to be more precise) with $w = t = \frac{n-k}{\log_2 n}$. This can be formulated as a new problem:

GOPPA PARAMETERIZED SYNDROME DECODING (GPSD):
  *Input:* a binary matrix $\mathcal{H}$ of size $2^m \times r$ and a syndrome $\mathcal{S}$ of size $r$.
  *Property:* there exists a word $y$ of weight $\leq \frac{r}{m}$ such that $y\mathcal{H} = \mathcal{S}$.

This new problem is NP-complete and this can be proven by reduction to THREE DIMENSIONAL MATCHING using the matrix presented in Fig. 2. The matrix $\mathcal{A}$ is simply padded with zeroes so as to get the proportions of a Goppa code correcting up to $|T|$ errors, that is, $n'$ and $r'$ must verify $|T| = \frac{3|T|+r'}{\log_2(n'+|U|)}$.

Suppose one has an algorithm able to solve any instance of GPSD and he feeds it the inputs $\mathcal{H}$ (defined as in Fig. 2) and $\mathcal{S} = (1, \ldots, 1, 0, \ldots, 0)$ containing $3|T|$ ones followed by $r'$ zeroes. The sum of $|T|$ rows of $\mathcal{H}$ will have a weight $\leq 3|T|$ on the first $3|T|$ coordinates. The only way to have the equality is to select $|T|$ rows among the $|U|$ top rows of $\mathcal{H}$ and 0 in the bottom part. A solution to our instance of GPSD is thus a solution to the underlying THREE DIMENSIONAL MATCHING problem.

We have the required reduction, the last constraint being that this reduction must be polynomial. For this we can set $n' = 2^{\lceil \log_2 |U| \rceil} - |U|$ so that $n' \leq |U|$ and $n' + |U|$ is the smallest power of 2 larger than $|U|$. Then $r' = |T| \times (\lceil \log_2 |U| \rceil - 3)$. This means that the proposed reduction is valid and polynomial as long as $|U| \geq 8$. The cases where $|U| < 8$ can anyway be solved in polynomial time and are not an issue. We can thus affirm that the GOPPA PARAMETERIZED SYNDROME DECODING problem is NP-complete.

As in the original 1978 paper, we can now try to deal with the problem of low weight codewords: the SUBSPACE WEIGHT problem. With the constraint induced by Goppa codes it can be formulated as:

GOPPA PARAMETERIZED SUBSPACE WEIGHT (GPSW):
  *Input:* a binary matrix $\mathcal{H}$ of size $2^m \times r$.
  *Property:* there exists a word $y$ of weight $2\frac{r}{m} + 1$ such that $y\mathcal{H} = 0$.

Note that here the constraint has been modified a little: a Goppa code correcting $t$ errors has a minimal distance of $2t + 1$, thus the words of minimal weight have a weight $2\frac{r}{m} + 1$. To prove that this problem is NP-complete, the first idea would be to simply combine the two constructions of Fig. 1 and Fig. 2 and pad matrix $\mathcal{B}$ with zeroes. This could work perfectly if the expansion from matrix $\mathcal{A}$ to matrix $\mathcal{B}$ was not so huge. Matrix $\mathcal{B}$ has a width of $3|T|(|U|+1)$ and requires to find a word of weight $3|T|^2 + 4|T|$, which when plugged into the Goppa constraint gives $3|T|^2+4|T| = 2\frac{r}{m}+1$, with $r \geq 3|T|(|U|+1)$. Thus, $m \geq \frac{6(|U|+1)}{3|T|+4}$ and the height $n = 2^m$ of matrix $\mathcal{H}$ is exponential in $|U|$: this reduction is not polynomial!

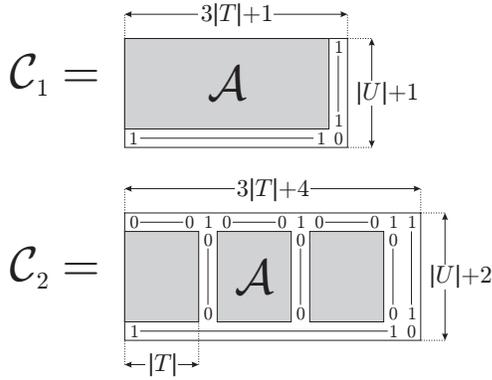

Fig. 3. More compact matrices used to reduce THREE DIMENSIONAL MATCHING to SUBSPACE WEIGHT.

We need to find another (more compact) construction for $\mathcal{B}$. For this we distinguish two case: if $|T|$ is even, use matrix $\mathcal{C}_1$ in Fig. 3, if $|T|$ is odd, use matrix $\mathcal{C}_2$. Suppose one has an algorithm solving any instance of SUBSPACE WEIGHT and wants to use it to solve an instance of THREE DIMENSIONAL MATCHING with $|T|$ even. He builds matrix $\mathcal{A}$ and expands it into matrix $\mathcal{C}_1$ of Fig. 3 by adding a row and a column of ones and a zero in the corner. Now $\mathcal{C}_1$ is input to the algorithm with $w = |T|+1$. A solution is a sum of $|T|+1$ rows summing to 0, thus, $|T|$ being even (and $|T|+1$ odd), this means that if a 0 is to appear in the last position, the bottom row has to be selected. The $|T|$ remaining lines are taken in $\mathcal{A}$ and must then sum to a row of ones. A solution is thus a solution to the THREE DIMENSIONAL MATCHING instance. Now if $|T|$ is odd, we simply add an element $\nu$ to $T$ and the vector $(\nu,\nu,\nu)$ to $U$ and can fall back to the case where $|T|$ is even. This translates in terms of incidence matrix in the matrix $\mathcal{C}_2$ of Fig. 3. Whatever the parity of $|T|$ we can thus solve any instance of THREE DIMENSIONAL MATCHING with a SUBSPACE WEIGHT algorithm. We have a valid reduction.

Now we can prove that GPSW is NP-complete by combining the constructions of Fig. 2 and Fig. 3. As for GPSD, we choose $n'$ such that $n$ is the smallest possible power of 2 greater than $|U|+1$ (or $|U|+2$) and obtain a polynomial time reduction as long as $|U| \gtrsim 64$. The GPSW problem is NP-complete.

*B. The General Case*

The proof we used for the Goppa Code case can trivially be adapted to any other constraint, as long as the following conditions are verified:

- the reduction must be polynomial: the dimensions $n$ and $r$ of $\mathcal{H}$ must be polynomials in $|U|$ and $|T|$,
- parameters for which the reduction cannot apply must be bounded by a finite value: for example the case $|U| \leq 8$.

The general problems can be formulated as follows:

PARAMETERIZED SYNDROME DECODING (PSD):
  *Input:* two integers $r$ and $w$, a binary matrix $\mathcal{H}$ of size $n \times r$ with $n = f(r,w)$, and a syndrome $\mathcal{S}$ of size $r$.
  *Property:* there exists $y$ of weight $\leq w$ such that $y\mathcal{H} = \mathcal{S}$.

PARAMETERIZED SUBSPACE WEIGHT (PSW):
  *Input:* two integers $r$ and $w$, a binary matrix $\mathcal{H}$ of size $n \times r$ with $n = f(r,w)$.
  *Property:* there exists $y$ of weight $w$ such that $y\mathcal{H} = 0$.

Intuitively, these two problems should be NP-complete for functions $f$ such that $f(3|T|,|T|)$ is polynomial in $|T|$ (and $f(3|T|+1,|T|+1)$, $f(3|T|+2,|T|+2)$). However, if we choose $f(r,w) = 2w$ (that is, we are looking for words of weight half the length $n$), even though $f$ is polynomial, it is clear that the proposed reduction fails to solve any instance of THREE DIMENSIONAL MATCHING where $|U| > 2|T|$: one must always have $n \geq |U|$. Being polynomial is not enough.

*Proposition 1:* Let $t = |T|$ and $u = |U|$. If there exists a function $g$, two polynomials $P$ and $Q$ and an integer $\lambda$ such that for all $t > \lambda$ and all $u > \lambda$, $3t \leq g(t,u) \leq P(t,u)$ and $u \leq f(g(t,u),t) \leq Q(t,u)$, then, our reduction is valid and the PSD problem is NP-complete.

  *Proof:* The polynomials $P$ and $Q$ are here to ensure that our reduction is polynomial. The value $\lambda$ makes sure that the only values for which one of the two bounds is not met are upper bounded: asymptotically, the reduction fits all instances. Then, once the two bounds are respected, the reduction works.

In the case of Goppa codes we had $f(r,w) = 2^{\frac{r}{w}}$ and chose $g(t,u) = t \times \lceil \log_2 u \rceil$ and $\lambda = 8$. This way, if $u > \lambda$ we have $3t \leq g(t,u) \leq tu$ and $f(g(t,u),t) = 2^{\lceil \log_2 u \rceil}$, so that $u \leq f(g(t,u),t) \leq 2u$.

We can state a very similar (but far less easy to read) proposition for the PSW problem.

*Proposition 2:* Let $t = |T|$ and $u = |U|$. If there exists two functions $g$ and $g'$, two polynomials $P$ and $Q$ and an integer $\lambda$ such that for all even value of $t > \lambda$ and all $u > \lambda$, $3t+1 \leq g(t,u) \leq P(t,u)$ and $u+1 \leq f(g(t,u),t) \leq Q(t,u)$, and for all odd value of $t > \lambda$ and all $u > \lambda$, $3t+2 \leq g'(t,u) \leq P(t,u)$ and $u+2 \leq f(g'(t,u),t) \leq Q(t,u)$, then, our reduction is valid and the PSW problem is NP-complete.

III. CONCLUSION

In this article some new variants of the Syndrome Decoding problem, fitting the specific parameters of Goppa codes are proven to be NP-complete. For this a new, more compact, NP-completeness reduction for the SUBSPACE WEIGHT problem is given. This proof should easily be generalized to other constraints for other classes of codes, as long as the constraint verifies some given properties. I leave it to other authors, with other specific constraints, to check wether or not the generic reduction I presented still works for them.